# Improving Image Classification of Knee Radiographs: An Automated Image Labeling Approach

Jikai Zhang, Carlos Santos, Christine Park, Maciej A. Mazurowski, Roy Colglazier

Corresponding Author: Jikai Zhang (814-321-6968, jikai.zhang@duke.edu) MB

Department of Electrical and Computer Engineering, Duke University, Durham, NC, United States

Room 10070, 2424 Erwin Road, Durham, NC, 27705

Carlos Santos, MS

Wake Forest University, Winston-Salem, NC, 27109

Christine Park, MS

Department of Radiology, Duke University Medical Center, Durham, NC, United States

Maciej A. Mazurowski, PhD

Department of Radiology, Duke University Medical Center, Durham, NC, United States

Department of Electrical and Computer Engineering, Department of Biostatistics and Bioinformatics, Department of Computer Science, Duke University, Durham, NC, United States

Roy Colglazier, M.D.

Department of Radiology, Duke University Medical Center, Durham, NC, United States

## Abstract

Large numbers of radiographic images are available in knee radiology practices which could be used for training of deep learning models for diagnosis of knee abnormalities. However, those images do not typically contain readily available labels due to limitations of human annotations. The purpose of our study was to develop an automated labeling approach that improves the image classification model to distinguish normal knee images from those with abnormalities or prior arthroplasty. The automated labeler was trained on a small set of labeled data to automatically label a much larger set of unlabeled data, further improving the image classification performance for knee radiographic diagnosis. We developed our approach using 7,382 patients and validated it on a separate set of 637 patients. The final image classification model, trained using both manually labeled and pseudo-labeled data, had the higher weighted average AUC (WAUC: 0.903) value and higher AUC-ROC values among all classes (*normal AUC-ROC*: 0.894; *abnormal AUC-ROC*: 0.896, *arthroplasty AUC-ROC*: 0.990) compared to the baseline model (WAUC=0.857; *normal AUC-ROC*: 0.842; *abnormal AUC-ROC*: 0.848, *arthroplasty AUC-ROC*: 0.987), trained using only manually labeled data. DeLong tests show that the improvement is significant on *normal* (p-value<0.002) and *abnormal* (p-value<0.001) images. Our findings demonstrated that the proposed automated labeling approach significantly improves the performance of image classification for radiographic knee diagnosis, allowing for facilitating patient care and curation of large knee datasets.

# Introduction

In modern radiology practices, large numbers of radiographic images are readily available for data-driven research in radiology[1–3]. These images are annotated for diagnosis of a wide range of pathologies and utilized by radiologists, orthopedics and other advanced practitioners toward guiding patient management and improving patient care for knee abnormality diagnosis. However, structured annotations for a large volume of radiographic images are difficult to obtain because the manual annotation process requires tremendous amounts of experts' attention and is very costly.

Deep learning (DL) solutions are typically developed using large numbers of labeled data. Specifically, image classification is an important supervised DL task to achieve various objectives in radiology, including disease detection, characterization, and monitoring[4–7]. An important example approached in this paper is classification of knee radiographs, which are commonly utilized for clinical evaluation of knee abnormalities[8–13] and knee arthroplasty[14–16]. Multiple studies incorporated DL-based image classification approaches for automating such evaluations[17–26]. These studies demonstrated the feasibility of leveraging large datasets for DL-based image classification in knee radiology, but a significant amount of effort in annotating knee radiographs has already been made to curate annotations for the existing datasets.

Alternatively, a label of a knee radiographic image can be directly interpreted and extracted from the corresponding radiology report. However, this is challenging for the following reasons. First, precise information extraction from unstructured knee radiology reports is difficult due to lack of standardization reporting[27]. Second, radiology reports are complex given the intricacy of knee pathologies in general, and reporting discrepancies or even errors across different radiologists[28]. Third, there is a lack of automated systems to interpret unstructured reports in knee radiology with high accuracy and sufficiency.

To overcome these limitations, this study aims to develop a hybrid DL framework, combining image classification and natural language processing (NLP) approaches, that leverages a large volume of unlabeled data to improve the performance of a multi-class classification model using only a limited number of manually labeled data. We decided on three classification labels, namely *normal*, *abnormal*, and *arthroplasty*. These labels represent visible conditions in the Bilateral posterior to anterior standing (BLPA) knee weightbearing views, which are commonly utilized to assess radiographic changes in the knee[8,29,30].

To the best of our knowledge, our proposed approach is the first to simultaneously utilize radiographs and their corresponding radiology reports in a DL framework for knee radiograph classification. The inherent association between a radiograph and its corresponding report ensures the same label for these two data sources. Our proposed framework consists of two main DL components. First, we developed an NLP-based automated labeler on labeled radiology reports and applied the labeler to a large volume of unlabeled reports to generate pseudo labels. Second, to validate the improvement when trained with additional pseudo-labeled data, we developed two pre-trained image classification models, one with only manually-labeled data and the other with additional pseudo-labeled data, and evaluated their performance on a hold-out test set.

# Methods

## Data Collection

This study was HIPAA-compliant and approved by the institutional review board (IRB) of <name withheld>. In the initial cohort, we retrospectively collected a dataset of 25,657 patients who had knee X-ray imaging studies finalized in 2019 within our large institutional health system. For each study, we downloaded radiology reports and knee radiographs in Digital Imaging and Communications in Medicine (DICOM) format from our electronic medical record (EMR) database.

We identified the initial cohort and obtained the dataset in two steps. First, we utilized a reporting workbench tool to search for radiology reports of knee X-ray imaging studies in our health system. The searching query was built based on the 10 most common knee procedure names in 2019 across our institution. The main searching results included (1) unique patient identifiers, (2) unique imaging study accession numbers, and (3) radiology reports. Second, we queried our large institutional picture archiving and communication system (PACS) to retrieve the DICOM objects for each imaging study by using the study accession numbers obtained from the main search.

In the final dataset, we only included patients with BLPA X-rays by filtering two DICOM attributes (1) modality= {CR, DX}, (2) Series Description= {PA Axial, PA Weight Bearing, PA Tunnel}. The final dataset consisted of 8,140 patients with 8,659 imaging studies. Each imaging study contained one radiology report and one BLPA radiograph.

## Annotation

BLPA radiographs were annotated with three labels: *normal, abnormal, and arthroplasty*, by using the handcraft rules, created by a team including four personnel: an expert radiologist and three non-expert researchers. The annotation rules consisted of descriptions for one category of the presence of arthroplasty, and eleven categories of knee abnormalities. Eleven abnormal categories included degenerative changes, postoperative changes or presence of non-arthroplasty orthopedic hardware, fractures, lesions, fragmentation, bone lucency, malalignment, osseous abnormalities, soft tissue abnormalities, developmental abnormalities, and trauma. The annotation team determined categories of images by applying the annotation rules to the associated radiology reports of BLPA radiographic images. A radiographic image was labeled *abnormal* if it met with at least one abnormal category, *arthroplasty* if it met with the arthroplasty category, and *normal* if none of *abnormal* or *arthroplasty* rules were met.

## Dataset Split

All patients (N=8,140) who had BLPA radiographs in 2019 were split into three groups based on the dates of radiology studies: (1) January to February, (2) March to November, and (3) December (Figure 1). Patients from January to February, with reports being manually labeled, were randomly split into the primary training set (TRAIN_PRI), the validation set for tuning hyperparameters and saving the best checkpoints (VAL_EVAL), and a held-out validation set for determining the final model (VAL_PTEST). Patients from March to November served as the secondary training set (TRAIN_SEC). Reports in TRAIN_SEC were not manually labeled, but were pseudo labeled by the NLP model. We removed 26 overlapping patients with those in the two validation sets to prevent "data leakage" issues. We called the above four dataset as the "development set". Patients from December were served as the test set (TEST) with all reports being manually labeled. Likewise, to prevent data leakage, we removed 44 overlapping patients from January to November, which had already been used in training and developing models.

## Automated Labeling of Reports Using an NLP model

The NLP model took a preprocessed report as an input and returned predictive probabilities of the three labels as an output. The final label had the highest predictive probability. Reports were preprocessed in two steps: (1) punctuations and numbers were removed and (2) only findings and impressions sections of the radiology reports were extracted. The model structure consisted of a feature extraction backbone and a classification module. An input unstructured report was tokenized and fed into the feature extraction backbone. We picked Bidirectional Encoder Representations from Transformers for Biomedical Text Mining (BioBERT)[31] model as the multi-class text classification backbone to output feature vectors, which were then fed into a linear layer with three output units for classification.

We trained candidate NLP models on TRAIN_PRI. The maximum token length of each report input to the model was fixed at 512. Batch size was set as 16, which is the optimal size for fitting the GPU RAM. Patience was set as 15 for early stopping purposes. Models were trained with the ADAM optimizer and tuned the following learning rates: {1e-5, 5e-5}. The model with the best performance on VAL_EVAL were used to annotate unlabeled reports in TRAIN_SEC. The NLP models were trained with PyTorch and HuggingFace[32].

## Image Classification Model

The image classification model took pixel arrays of an image as an input and returned predictive probabilities of the three labels. The final label had the highest predictive probability. We extracted pixel arrays from each corresponding DICOM file of an image and normalized them into the [0, 255] range. Similar to the NLP model structure, the imaging model structure consisted of a feature extraction backbone and a classification module. In this work, we picked EfficientNet-b4[33] as the multi-class image classification backbone output feature vectors, which were then fed into a linear layer with three output units for classification.

The baseline image classification model used manual labels and was trained on TRAIN_PRI. Our proposed model leveraged additional pseudo-labeled data in TRAIN_SEC. The input image size for training was 380*380, the same as the size used to train a full EfficientNet-b4 model in the original paper. Batch size was set as 16, which is the optimal size given the GPU RAM. Patience was set as 10 for early stopping purposes. We trained image classification models using the ADAM optimizer and tuned the following learning rates: {1e-5, 5e-5}. We determined the final learning rate configuration by selecting the model with the best performance on VAL_EVAL. Using this configuration, we evaluated performance of models with five random experiments on VAL_PTEST and selected the best seed to be tested on TEST. Image classification models were trained using PyTorch with the "EfficientNet PyTorch" package.

## Transfer Learning

Training feature extraction backbones, namely BioBERT and EfficientNet-b4, from scratch required significant amount of data and computational resources. Thus, to expedite the training while maintaining high accuracy, we utilized "transfer learning" approach for training our classification models. To do so, we initialized weights of feature extraction backbones using pre-trained weights and further fine-tuned such model on the downstream classification task. Weights of BioBERT were initialized using pre-trained BioBERT-Base v1.1 weights optimized on English Wikipedia, BooksCorpus, PubMed abstracts, and fine-tuned in downstream training; weights of EfficientNet-b4 were initialized using pre-trained EfficientNet weights optimized on ImageNet.

## Statistical Analysis

Each model was trained with five random seeds. We collected mean, standard deviation, and median of weighted area under receiver operating characteristic curve (WAUC) based on five experiments of each model. WAUC-ROC is calculated as following:

$$WAUC = f_1 AUC_1 + f_2 AUC_2 + f_3 AUC_3$$

, where $f_i$ and $AUC_i$ is the frequency percentage and AUC of label $i$ in TEST respectively. When we compared two models, the model with larger mean WAUC was deemed as a better model and selected as the representative for further evaluations. Once we obtained the final model, we reported the AUC for each class and WAUC, and plotted ROC curves for each class accordingly. The DeLong test[34] was performed on each class to evaluate if there was a significant difference in AUC-ROC when adding pseudo-labeled data in training.

# Results

### Dataset Analysis
A total of 8,019 patients (7,382 in the development set that includes TRAIN_PRI, VAL_EVAL, VAL_PTEST, TRAIN_SEC, 637 in the test set) and 8,585 images were included in the final analysis. In the development set, the mean age was 53 years old with a standard deviation (SD) of 19 years old. The youngest patient was 2 years old and the oldest patient was 96 years old. 2988 (40%) patients were male and 4394 (60%) patients were female. 4899 (66%) patients were white, 1694 (23%) patients were black, 182 (3%) patients were Asian. In the test set, the mean age was 54 years old with SD of 19 years old. The youngest patient was 6 years old and the oldest patient was 97 years old. 276 (43%) patients were male and 361 (57%) patients were female. 428 (67%) patients were white, 133 (21%) patients were white, and 24 (3%) patients were Asian (Table 1).

TRAIN_PRI contained 854 cases, with 198 (23%) normal cases. VAL_EVAL and VAL_PTEST contained 208 cases each, with 41 (20%) and 44 (21%) normal cases respectively. TRAIN_SEC contained 6656 cases, with 1867 (28%) cases being pseudo-labeled normal. TEST contained 659 cases, with 151 (23%) normal cases (Table 2). The ratio of manually-labeled data to pseudo-labeled data used in the development set was approximately 1:5.2.

### Test results of BLPA image classification
The NLP model had the best performance of WAUC=0.995 (AUC_normal=0.993, AUC_abnormal=0.996, AUC_arthroplasty=1.000) on VAL_EVAL, indicating near-perfect quality of pseudo labels. Table 3 shows the results on TEST using only manually labeled data vs. manually and automatically labeled data. When training with additional pseudo-labeled data, all reported metrics were higher (AUC-ROC Normal + 0.052, AUC-ROC Abnormal +0.048, AUC-ROC Arthroplasty +0.003, WAUC +0.046) than training with only manual-labeled data. The model almost perfectly predicted images with evidence of arthroplasty hardware in both specifications (Manual-labeled AUC-ROC Arthroplasty = 0.987, Manual + Pseudo-labeled AUC-ROC Arthroplasty = 0.990). For each class, we plotted ROC curves in Figure 2 and performed the DeLong test. Results from the DeLong test showed that there is significant improvement in Normal AUC-ROC (p-value<0.002) and Abnormal AUC-ROC (p-value<0.001).

# Discussion

In this study, we proposed a DL framework that improves the baseline image classification performance by introducing a large volume of unlabeled data. A state-of-the-art NLP model, trained on a small set of labeled reports, served as an automated labeler to provide accurate pseudo-labels of unlabeled reports and their corresponding images. By augmenting the training size by approximately eight times, we trained image classification models with additional pseudo-labeled images and achieved significantly better classification results on *normal* (WAUC+0.052, p-value_normal<0.002) and *abnormal* (WAUC+0.048, p-value_abnormal<0.001) images. Although no significant improvements were found for

*arthroplasty* images, models had already achieved high performance in both settings (AUC=0.987 using only manually-labeled cases; AUC=0.990 using additional pseudo-labeled cases) and improved only marginally when we added pseudo-labeled cases (WAUC+0.003).

Our findings demonstrate important practical values. First, by applying the DL framework, our approach requires only 16% of the development set to be manually annotated. Such low percentage of required annotations significantly reduced the burden of human annotations and improved the inherit limitations of human interpretation, such as observer variability, time constraints, cost and bias. Second, similar to the literature in which BioBERT and EfficientNet models have been proven to be effective feature extraction backbones in general radiology text[35,36] and image classification[37–39] tasks, our results demonstrated the power and potentials of DL tools in knee radiology. When our framework is adopted to other datasets in future research, the DL backbones can be easily substituted with other tools for optimal usage. Third, our dataset split plan reflected a real-world scenario where DL developers trained models using retrospective data and validated performance on prospective data. Hence, we believe that our approach can be adopted to provide reliable assistance in the clinical applications related to the identification and diagnosis of knee abnormalities using knee radiographs.

Pseudo-labeling is an important feature of our framework. The idea of pseudo-labeling is fundamental in conventional SSL tasks [40–43]. Such tasks have also been proven to be effective in predicting knee abnormalities[44,45]. SSL-based algorithms require a small amount of labeled input, train the whole model with a joint input of labeled and unlabeled cases with, and learn the pseudo labels for unlabeled cases accordingly. Our approach shares the same goal with SSL tasks: improve the classification performance by utilizing additional pseudo-labeled cases generated from a large unlabeled dataset. However, unlike SSL tasks, first, we utilized NLP to learn knowledge from the manually labeled reports to automatically pseudo-label the unlabeled reports. Thanks to the inherent association between radiology reports and images, we treated report labels as image labels. Second, in a separate model, we trained both labeled and pseudo-labeled images for final prediction. We acknowledge that such association might be vulnerable in practice. Thus, future work can involve SSL-based models to learn pseudo-labels on the same input source, without considering the inherent link between different types of input sources.

Another important feature of our framework is to exploit both radiology reports and images in one workflow for image classification using a large volume of unlabeled radiology data. In the literature, multiple studies have shown the effectiveness of including this feature for curating large datasets of radiographs. For example, Jeremy et al curated the CheXpert dataset and investigated different approaches of incorporating uncertainty labels in training to predict lung pathologies[46]. Xiaosong et al curated the ChestX-ray8 dataset aided by a concept detection tool and proposed an image classification model to detect and locate thoracic disease[47]. In both studies, curation of large datasets was aided by an automated NLP labeler. The curated data were then tested feasible for downstream image classification tasks. While our proposed framework shared such feature, a key difference is that we utilized a state-of-the-art DL-based NLP model as the automated labeler to generate high-quality pseudo-labels for knee radiology reports. This demonstrates great potentials in applications of our framework to curate large datasets of knee radiographs. Future work can focus on applying our framework to significantly larger datasets of unlabeled knee radiographs and further validated for downstream tasks.

## Limitations

We acknowledge several limitations in this study. First, our proposed framework was trained and validated using data in one health system without being externally validated. Future work could apply our framework to multicenter data and test the generalizability of our approach. Second, annotation rules were developed based on the experience of a single expert. As a result, our defined rules may not comprehensively reflect all characteristics on knee radiographs. Third, label noises may exist in pseudo-labeled cases that were generated by the trained NLP model. However, we believe that the image classification model can tolerate such label noise because of the high-quality pseudo-labels. Fourth, we determined the selection criteria of BLPA view by manually reviewing a small number of knee radiographs in the dataset. Future work could involve a more rigorous process in consolidating such criteria. Lastly, our utilization of BLPA view for this study does not account for the differences in joint space loss that can be seen with variable flexion of the knee which has been shown in other studies[8].

## Conclusion

By harnessing DL powers for annotating a large volume of unlabeled reports using only a small number of labeled data, we have shown the feasibility of our proposed approach to improve image classification performance for knee radiographic diagnosis, without the labor of interpreting an overwhelming number of images. As a result, the proposed approach minimizes the inherent limitations of human annotations and can be potentially useful for curating large knee datasets. Our focus on a commonly used imaging modality of the knee will allow for wide utilization of the application in knee radiology to improve patient care for knee abnormality diagnosis.

# Tables

Table 1. Demographic characteristics of the patients at baseline

| Characteristic | Development Set (N=7382) | Test set (N=637) |
|---|---|---|
| Mean Age ± SD* year (range) | 53 ± 19 (2 – 96) | 54 ± 19 (6 – 97) |
| Male No. (%) - Female No. (%) | 2988 (40%) - 4394 (60%) | 276 (43%) – 361 (57%) |
| Race or ethnic group No. (%) | | |
|   White | 4899 (66%) | 428 (67%) |
|   Black | 1694 (23%) | 133 (21%) |
|   Asian | 182 (3%) | 22 (3%) |
|   Other** | 607 (8%) | 54 (9%) |

* SD: Standard Deviation
** Category "Other" includes American Indian or Alaska Native, Native Hawaiian or other Pacific Islander, not reported/unavailable, not Hispanic or Latino.

Table 2: Label distributions of images in January to February, including TRAIN_PRI, VAL_EVAL, and VAL_PTEST.

| Dataset | Normal No. (%) | Abnormal No. (%) | Arthroplasty No. (%) | Total No. |
|---|---|---|---|---|
| TRAIN_PRI | 198 (23%) | 589 (69%) | 67 (8%) | 854 |
| VAL_EVAL | 41 (20%) | 152 (73%) | 15 (7%) | 208 |
| VAL_PTEST | 44 (21%) | 142 (68%) | 22 (11%) | 208 |
| TRAIN_SEC* | 1867 (28%) | 4282 (64%) | 507 (8%) | 6656 |
| TEST | 151 (23%) | 457 (69%) | 51 (8%) | 659 |

* Labels in TRAIN_SEC were pseudo-labeled by the NLP model

Table 3. Comparison of performance on TEST using different data based on the configuration selection on VAL_PTEST. AUC-ROC for each class and WAUC are provided

| Data used for training | AUC-ROC Normal | AUC-ROC Abnormal | AUC-ROC Arthroplasty | WAUC |
|---|---|---|---|---|
| Only manually labeled data (TRAIN_PRI) | 0.842 | 0.848 | 0.987 | 0.857 |
| Manually and automatically labeled data (TRAIN_PRI + pseudo-labeled TRAIN_SEC) | 0.894 | 0.896 | 0.990 | 0.903 |

# Figure Captions

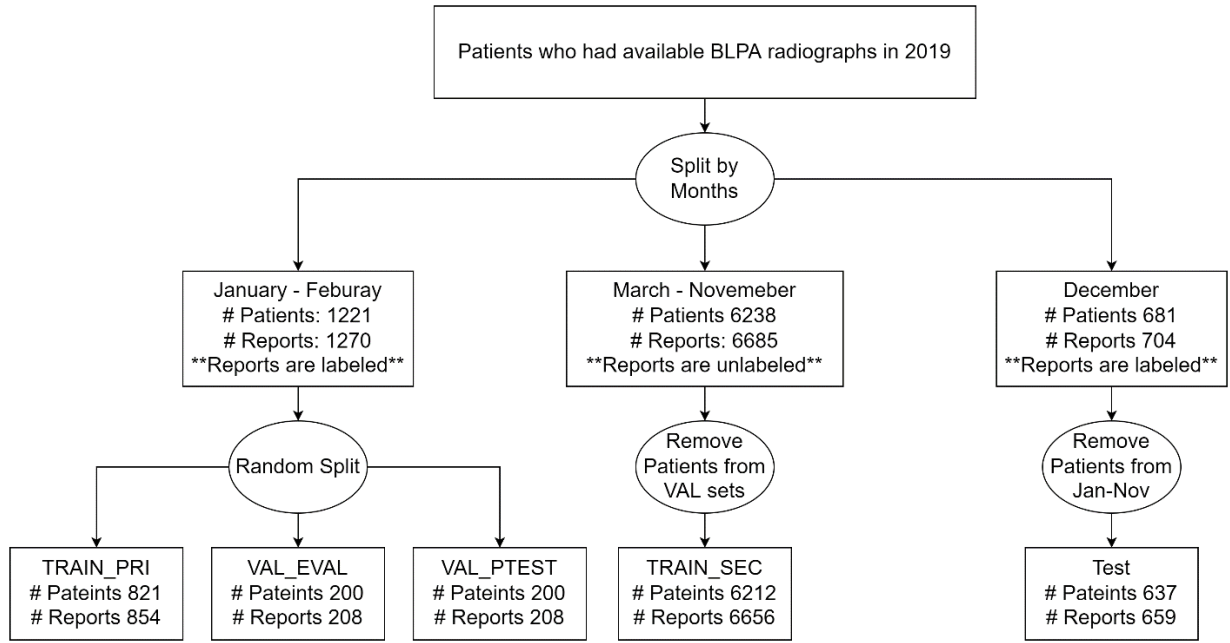

Figure 1: Dataset Split Plan. *TRAIN_PRI:* used for training both image/NLP. *VAL_EVAL:* used for saving checkpoints/tuning hyperparameters during training. *VAL_PTEST:* used for last evaluations before applying models to *TEST*. *TRAIN_SEC*: used for automatic pseudo labeling. We called the above 4 dataset as the "development set". *TEST:* A completely untouched hold-out set for final evaluation.

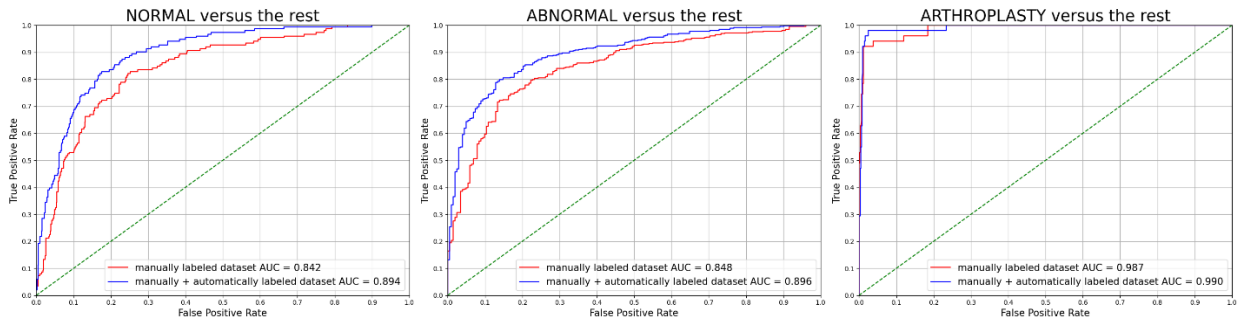

Figure 2. ROC plots on *TEST* of the specification with the best median AUC-ROC on VAL_EVAL in each tested model using different data: red: Only manually labeled data (only TRAIN_PRI, AUC-ROC = 0.857), blue: Manually and automatically labeled data (TRAIN_PRI + pseudo-labeled TRAIN_SEC, Average AUC-ROC = 0.903). DeLong tests provide p-values for each class: p-value_normal<0.002, p-value_abnormal<0.001, p-value_arthroplasty = 0.4006.